\documentclass[aps,prd,nofootinbib,reprint,superscriptaddress,longbibliography]{revtex4-1}
\usepackage{etoolbox}
\usepackage{dcolumn}
\usepackage{bm}
\usepackage{graphics}
\usepackage{dirtytalk}
\usepackage{graphicx} 
\usepackage{blindtext}
\usepackage{amsmath}
\usepackage{amsfonts}
\usepackage{amssymb}
\usepackage[%
colorlinks=true,
urlcolor=blue,
linkcolor=red,
citecolor=blue
]{hyperref}

\usepackage{wrapfig}

\usepackage{natbib}
\begin{document}
\title{\bf The role of mutual information in the Page curve}
\vskip 1cm
	\vskip 1cm
\author{Anirban Roy Chowdhury}
\email{iamanirban.rkmvc@gmail.com}
\affiliation{Department of Astrophysics and High Energy physics\linebreak
	S.N.~Bose National Centre for Basic Sciences, JD Block, Sector-III, Salt Lake, Kolkata 700106, India}	
\author{Ashis Saha}
\email{ashisphys18@klyuniv.ac.in}
\affiliation{Department of Physics, University of Kalyani, Kalyani 741235, India}
\author{Sunandan Gangopadhyay}
\email{sunandan.gangopadhyay@bose.res.in}
\affiliation{Department of Astrophysics and High Energy physics\linebreak
	S.N.~Bose National Centre for Basic Sciences, JD Block, Sector-III, Salt Lake, Kolkata 700106, India}
	\begin{abstract}
	\noindent In this work, we give two proposals regarding the status of connectivity of entanglement wedges and the associated saturation of mutual information. The first proposal has been given for the scenario before the Page time depicting the fact that at a particular value of the observer's time $t_b=t_R$ (where $t_R\ll\beta$), the mutual information $I(R_+:R_-)$ vanishes representing the disconnected phase of the radiation entanglement wedge. We argue that this time is the Hartman-Maldacena time at which the fine-grained entropy of radiation goes as $S(R)\sim \log(\beta)$, where $\beta$ is the inverse of Hawking temperature of the black hole. On the other hand, the second proposal probes the crucial role played by the mutual information of black hole subsystems in obtaining the correct Page curve of radiation.
	\end{abstract}
\maketitle
\noindent 
The occurrence of pair production that occurs in the near-horizon region of the black hole has led to one of the most remarkable and mysterious phenomenon in theoretical physics, that of Hawking radiation \cite{Hawking:1975vcx}. In the context of contemporary theoretical physics, this effect has been a matter of great interest. This is due to the fact that being a quantum mechanical radiation, its existence is a direct signature of the microscopic physics involved in the theory of general relativity. This in turn has motivated researchers to study its quantum mechanical aspects, such as von Neumann entropy \cite{Chuang:2000}. However, the study of von Neumann entropy of Hawking radiation \cite{PhysRevD.14.2460} has in turn gifted us an elegant paradox. The paradox can be put forward in the following way. It has been observed that the formation of a black hole (due to the gravitational collapse of a massive shell) is associated with a pure state with the von Neumann entropy of radiation being zero. Further, the concept of unitary evolution demands that the final state (at the end of the evaporation process) should also be a pure state and hence the von Neumann entropy should again vanish. However, Hawking's semi-classical analysis has showed that for an evaporating black hole, the von Neumann entropy of Hawking radiation is an ever increasing quantity with respect to the time of the observer \cite{PhysRevD.14.2460} and it does not vanish after the evaporation of the black hole. This observation is in conflict with Page's observation. It is also to be noted that the von Neumann entropy of the radiation is identified with the von Neumann entropy of matter fields located on the region $R$ outside the black hole. Furthermore, it is also to kept in mind that the state on the full Cauchy slice is a pure one and therefore we can write $S_{vN}(R)=S_{vN}(R^c)$. So a natural question arises regarding the correct time evolution of the von Neumann entropy of Hawking radiation. This was efficiently addressed by the so-called \textit{Page curve}. The Page curve curve suggested that in order to satisfy the unitarity condition, the von Neumann entropy of the radiation shall start from zero and monotonically increase upto the \textit{Page time} and then again drop down to zero, signifying the end of the evaporation process \cite{PhysRevLett.71.3743,Page_2013}. The Page curve has also simplified the scenario by pointing out that the paradox only appears after the Page time. The reason for this is that in this region, the von Neumann entropy of radiation $S_{vN}(R)$ is greater than the coarse grained entropy (Bekenstein-Hawking entropy \cite{Bekenstein:1972tm,Bekenstein:1973ur,PhysRevD.13.191}) $S_{BH}$ of the black hole, that is $S_{vN}(R)> S_{BH}$. Keeping in mind the unitarity evolution of radiation, many interesting approaches has been made to address this issue \cite{Almheiri:2012rt,Almheiri:2013hfa,Lloyd:2013bza,Papadodimas:2013wnh}.  It was also suggested that the exact Page curve can only be obtained from the full quantum theory of gravity. Some recent works in this direction can be found \cite{Yu:2021cgi,Ahn:2021chg,Krishnan:2020fer,Krishnan:2020oun}.\\
In recent times, the concept of entanglement wedge reconstruction from Hawking radiation has suggested that certain regions in the interior of black hole contributes to the fine grained entropy of Hawking radiation \cite{Penington:2019npb,Almheiri:2019psf,Almheiri:2019hni,Almheiri:2019yqk}. These auxiliary regions are denoted as the \textit{islands} together with their end points identified as the \textit{quantum extremal surfaces} (QES) \cite{Engelhardt:2014gca,Engelhardt:2019hmr,Akers:2019lzs,Wall:2012uf}. It is to be mentioned that the quantum extremal surfaces are the quantum corrected classical extremal surfaces \cite{PhysRevLett.96.181602,Hubeny:2007xt}. On the other hand, the fine-grained entropy of radiation is the generalized version of the von Neumann entropy. From the holographic perspective, the generalization represents a bulk (area) term together with the quantum corrected extremal surface term \cite{Faulkner:2013ana}. In the presence of the island in the black hole interior, the fine-grained entropy of the Hawking radiation is given by
\begin{equation}\label{eq1}
 S(R) =	\textrm{min}~\mathop{\textrm{ext}}_{\mathcal{\mathrm{I}}}\bigg\{\frac{\textrm{Area}(\partial I)}{4G_N}+S_{vN}(I\cup R)\bigg\}~.
\end{equation}
From the semi-classical point of view, the islands originate from the replica wormhole saddle points (with proper boundary conditions) of the gravitational path integral, which appears due to the application of replica technique in dynamical gravitational background \cite{Almheiri:2019qdq,Penington:2019kki,Goto:2020wnk,Colin-Ellerin:2020mva}. Due to this remarkable observation, the island formulation has emerged as an important prescription to be studied \cite{Almheiri:2020cfm,chen2020information,Hashimoto:2020cas,Hartman:2020swn,Anegawa:2020ezn,Dong:2020uxp,Balasubramanian:2020xqf,Raju:2020smc,Alishahiha:2020qza,Azarnia:2021uch,Arefeva:2021kfx,He:2021mst,Omidi:2021opl,Yu:2021rfg,Yadav:2022fmo,Du:2022vvg}.\\
In \cite{Saha:2021ohr}, the role played by the mutual information has been probed in the context of the fine grained entropy of Hawking radiation of BTZ black hole. It was shown that the condition of vanishing mutual information between the subsystems leads to a time independent expression for the fine grained entropy of the Hawking radiation which is consistent with the correct Page curve. 
In this work, we present some more new insights and proposals.
\section{Gravitational set up: JT gravity + flat baths}
We begin our analysis by considering the two dimensional eternal black hole solution of Jackiw-Teitelboim (JT) gravity \cite{Teitelboim:1983ux,Jackiw:1984je}, coupled to a pair of non-gravitational auxiliary thermal baths (flat spacetimes). However, we will show that our observation can be applied in any spacetime dimensions under certain simple approximations. JT gravity is one of many solutions of the two dimensional dilaton gravity theory. The action for this theory reads \cite{Mandal:1991tz,Grumiller:2002nm}
\begin{eqnarray}\label{eq2}
I_{2d} &=&\frac{1}{16\pi G_N}	\int_{V}^{} \sqrt{-g}\left[\phi R+U(\phi)(\nabla\phi)^2+V(\phi)\right]d^2x\nonumber\\
&&+\frac{1}{8\pi G_N}\int_{\partial V}\sqrt{-h} \phi K dx~.
\end{eqnarray}
The JT gravity solution can be obtained by setting $U(\phi)=0$ and $V(\phi)=\frac{2\phi}{l^2}$, where $\phi$ represents the dilaton field profile $\phi(r)=\frac{r}{l}$. The above action also leads to the CGHS black hole \cite{Callan:1992rs} corresponding to different value of the functions $U(\phi)$ and $V(\phi)$ \cite{Grumiller:2002nm}. On the other hand, the JT gravity black hole solution of the above theory reads (in the Schwarzschild gauge)
\begin{eqnarray}\label{eq3}
ds^2 = -f(r)dt^2+\frac{dr^2}{f(r)};~f(r)=\frac{(r^2-r_+^2)}{l^2}~.	
\end{eqnarray} 
The corresponding Hawking temperature and coarse grained entropy of the black hole are obtained to be $T_H=\beta^{-1}=\frac{r_+}{2\pi l^2}$ and $S_{BH} = \frac{r_+}{4G_N l}$. We now assume that the above geometry and the coupled flat spacetimes (non-gravitational baths) are filled with $2d$ free CFT, for instance theory of free fermions (or other theories equivalent to free fermions). This procedure of gluing thermal baths with eternal $AdS$ black hole incorporates the necessary absorbing boundary conditions for the outgoing Hawking quanta \cite{Hemming:2000as,Almheiri:2013hfa,VanRaamsdonk:2013sza}. In Kruskal coordinates, the metric inside the $AdS$ spacetime reads
\begin{eqnarray}\label{eq4}
ds^{2}_{JT}=-F^{2}(r)dudv~;~F^{2}(r)=-\frac{f(r)}{\kappa^{2}uv}~.
\end{eqnarray}
It is to be noted that the above metric is valid only for the gravitational regions. However, as we have mentioned, the bath regions are flat and this in turn means that the metric corresponding to these regions should also carry this signature. In order to obtain this metric, we follow the approach given in \cite{Almheiri:2019yqk} and \cite{He:2021mst}. We assume that the curved spacetime influences of the JT gravity vanishes at a certain hypothetical cut-off distance, namely, $r_{R(L)}=\xi$, which lies inside the AdS boundary depicted by the vertical lines in Fig.\eqref{fig1} and Fig.\eqref{fig2}. 
Now in order to ensure that both of the metrics (associated to bath and JT gravity) are continuously connected at $r_{R(L)}=\xi$, we extend the Kruskal coordinates to the bath regions by incorporating the normalization condition of the \emph{tortoise coordinate} $\lim_{r\rightarrow\infty} r^*(r)=0$. These considerations help us to write down following flat form of the bath metric \cite{He:2021mst}
\begin{eqnarray}\label{bath metric}
ds^{2}_{Bath}=-F^{2}(\xi,r)dudv~;~F^{2}(\xi,r)=-\frac{f(\xi)}{\kappa^{2}uv}
\end{eqnarray}
where $\xi =\alpha r_+$ and $\alpha\gg 1$. Before we proceed to the subsequent analysis, we specify the Kruskal coordinates. For the right wedge (RW), this reads
\begin{eqnarray}
u&=&-e^{-\kappa(t-r^*(r))}\nonumber\\
v&=&e^{\kappa(t+r^*(r))}	
\end{eqnarray}
and for the left wedge (LW) reads
\begin{eqnarray}
	u&=&e^{\kappa(t+r^*(r))}\nonumber\\
	v&=&-e^{-\kappa(t-r^*(r))}	
\end{eqnarray}
where $\kappa=\frac{r_+}{l^2}$ is the surface gravity. 
\section{Before Page time scenario: probing the role of $I(R_+:R_-)$} 
\noindent In this work, we first discuss the before Page time ($t_p$) scenario. For $t_{obs}<t_p$, the Hawking saddle point of the gravitational path integral dominates and the fine grained entropy of the Hawking radiation is identified by the von Neumann entropy of the matter fields.
\begin{figure}[htb]
	\centering
	\includegraphics[scale=0.55]{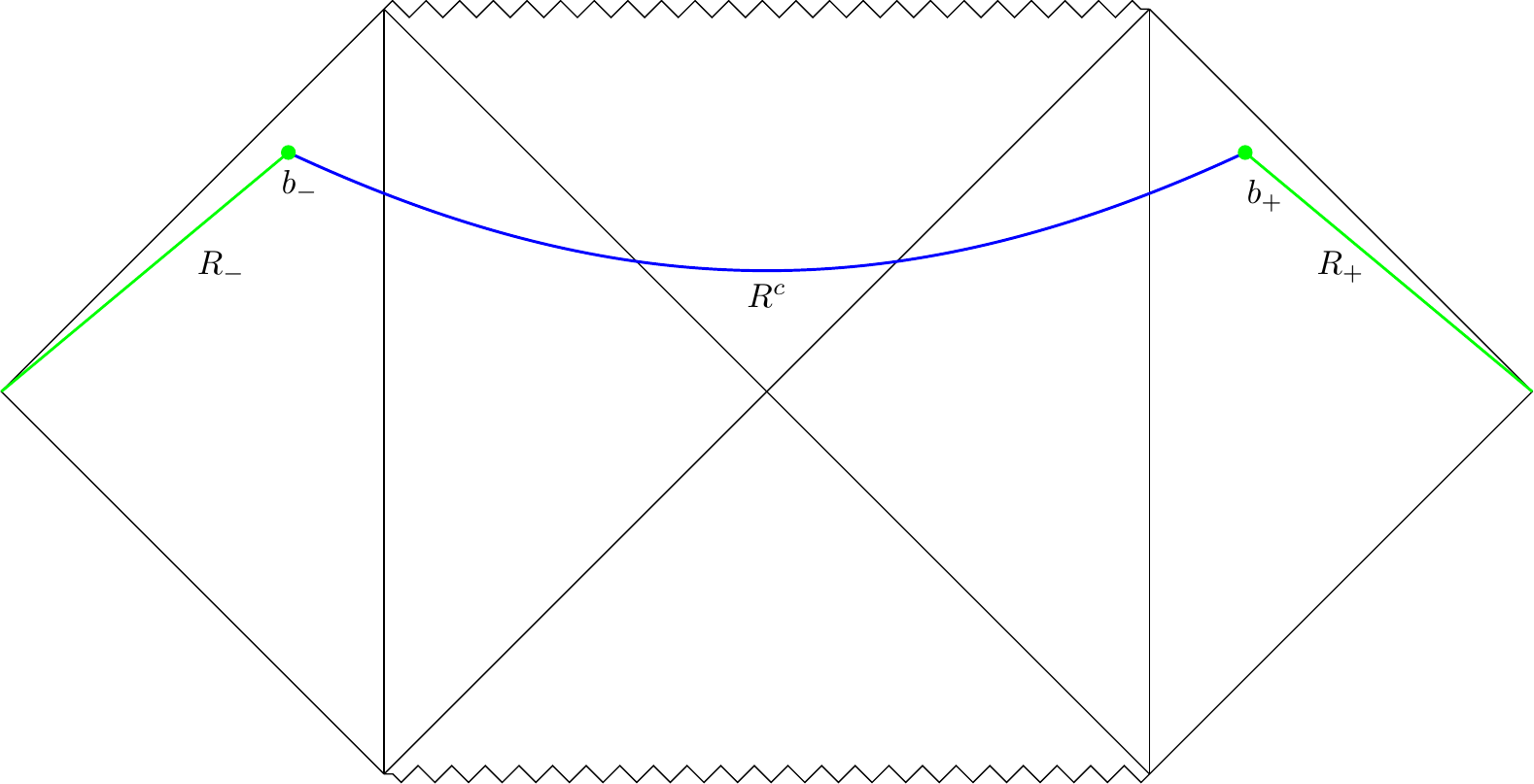}
	\caption{Penrose diagram of eternal black hole of JT gravity + flat auxiliary thermal bath system. The $R_{\pm}$ regions have been shown in green with the inner boundaries $b_{\pm}=(\pm t_b,b)$ \cite{He:2021mst}.}
	\label{fig1}
\end{figure}
This in turn means that one neglects the island contribution in eq.\eqref{eq1} and get $S(R)=S_{vN}(R)$. Now the region outside the black hole  comprises of two disjoint intervals, that is $R_+$ and $R_-$ (where the $\pm$ signifies the right and left wedges of the Pensrose-Carter diagram). The endpoints of the disjoint regions $R_{\pm}$ are $[e_{\pm}:b_{\pm}]$. As $R_{\pm}$ regions are extended to spatial infinity from the inner boundary $b_{\pm}=(\pm t_b,b)$, we introduce the point $e_{\pm}$ in order to regularize it, that is, $e_{\pm}=(0,e)$. We will eventually take $e\rightarrow\infty$. This has been shown graphically in the Penrose diagram given in Fig.\eqref{fig1}. In this set up, the fine-grained entropy of radiation reads
\begin{eqnarray}\label{eq5}
S_{vN}(R) = S_{vN}(R_+\cup R_-)~,~ R = R_{+}\cup R_{-}~.
\end{eqnarray}
Further, as the state of matter on the full Cauchy slice is a pure state, one can also write down the following
\begin{eqnarray}\label{eq8}
	S_{vN}(R_+\cup R_-)&=&S_{vN}(R^c)	
\end{eqnarray}
where $R^c$ is the complement region of $R=R_+\cup R_-$. This in turn means in order to quantify the entanglement entropy of matter fields (free $2d$ CFT) on $R$, we need to compute the following quantity
\begin{eqnarray}\label{sr}
S_{vN}(R^c)=\left(\frac{c}{3}\right)\log d(b_+,b_-)~.	
\end{eqnarray}
The distance $d(b_+,b_-)$, given in the above expression can be computed explicitly from the metric given in eq.\eqref{bath metric}. The expression of $d(b_{+},b_{-})$ is given by
\begin{eqnarray}
	d(b_{+},b_{-})=2F(\xi,b) e^{\kappa r^{*}(b)}\cosh(\kappa t_{b})~.
\end{eqnarray}
Now using the above expression in eq.(\ref{sr}), the entropy of Hawking radiation comes out to be
\begin{eqnarray}\label{Neweq1}
	S(R)&=&S_{vN}(R_+\cup R_-)\nonumber\\
 &=&\left(\frac{c}{3}\right)\log\left[\left(\frac{\beta}{2\pi}\right) \sqrt{f(\xi)}\cosh\left(\frac{2\pi t_b}{\beta}\right)\right]~.	
\end{eqnarray}
It can been observed from the above expression that, at the early times ($t_b\ll\beta$), the fine grained entropy (without the island contribution) of Hawking radiation $S(R)=S_{vN}(R_+\cup R_-)$ reduces to the following form
\begin{eqnarray}
S(R) &\approx& \left(\frac{c}{3}\right)\log\left[\left(\frac{\beta}{2\pi}\right) \sqrt{f(\xi)}\right]+\left(\frac{c}{6}\right)\left(\frac{2\pi t_b}{\beta}\right)^2~.\nonumber\\	
\end{eqnarray}
On the other hand, at late times ($t_{P}>t_b\gg\beta$), where $t_{P}$ is the Page time, one obtains the following form of $S_{vN}(R_+\cup R_-)$
\begin{eqnarray}
S(R) &\approx&	\left(\frac{c}{3}\right)\log\left[\left(\frac{\beta}{2\pi}\right) \sqrt{f(\xi)}\right]+\left(\frac{c}{3}\right)\left(\frac{2\pi t_b}{\beta}\right).\nonumber\\
\end{eqnarray}
This implies that as long as the replica wormhole saddle points (island) does not contribute, $S(R)$ increases in both early and late times, with respect to the observer's time $t_b$. However, the nature of increment in these two time domains are different as in the early times it increases as $\sim t_b^2$ and in the late times, it increases linearly ($\sim t_b$). This observation firmly agrees with the one shown in \cite{Hartman:2013qma}.\\
Further, one can also compute the von Neumann entropy of matter fields on the individual intervals $R_+$ and $R_-$. This can done by using the following expressions
\begin{eqnarray}\label{sr+}
	S_{vN}(R_+)&=&\left(\frac{c}{3}\right)\log d(b_+,e_+)\nonumber\\
	S_{vN}(R_-)&=&\left(\frac{c}{3}\right)\log d(b_-,e_-)~.
\end{eqnarray}
We can compute the distances by using the flat metric given in eq.(\ref{bath metric}). The expressions of $d(b_{+},e_{+})$ and $d(b_{-},e_{-})$ read
\begin{eqnarray}
	d(b_{+},e_{+})&=&\sqrt{2F(\xi,b)F(\xi,e)e^{\kappa r^{*}(b)}[\cosh(\kappa r^{*}(b))-\cosh(\kappa t_{b})]}\nonumber\\
	&=&d(b_{-},e_{-})~.
\end{eqnarray}
Substituting the above expression in eq.(\ref{sr+}), 
we have the following result
\begin{widetext}
	\begin{eqnarray}\label{Neweq2}
S_{vN}(R_+) = S_{vN}(R_-) =\left(\frac{c}{6}\right)	\log\left[2\left(\frac{\beta}{2\pi}\right)^2f(\xi)\left\{|\cosh\left(\frac{2\pi r^*(b)}{\beta}\right)-\cosh\left(\frac{2\pi t_b}{\beta}\right)|\right\}\right]~.	
	\end{eqnarray}	
\end{widetext}
From the expressions given in eq.\eqref{Neweq1} and eq.\eqref{Neweq2}, one can now compute the mutual information between the matter fields on the intervals $R_+$ and $R_-$. This reads
\begin{widetext}
\begin{eqnarray}\label{Newqeq3}
	I(R_+:R_-) &=& S_{vN}(R_+)+ S_{vN}(R_-)-S_{vN}(R_+\cup R_-)\nonumber\\
	&=&\left(\frac{c}{3}\right)\log\left[\left(\frac{\beta}{2\pi}\right)\sqrt{f(\xi)}\left\{\frac{|\cosh\left(\frac{2\pi r^*(b)}{\beta}\right)-\cosh\left(\frac{2\pi t_b}{\beta}\right)|}{\cosh\left(\frac{2\pi t_b}{\beta}\right)}\right\}\right]~.\nonumber\\
\end{eqnarray}
\end{widetext}
With the above expression in hand, we now look at the behaviour of $I(R_+:R_-)$ in both early ($t_b \ll \beta$) and late time  ($t_b \gg \beta$) domains. Before we proceed to do that, we compute the value of $I(R_+:R_-)$ at $t_b=0$. This is found to be $I(R_+:R_-)|_{t_b=0}\approx 1.17$, where we choose $l=10r_+, b\sim\alpha = 15r_+$ and $r_+=10$. This in turn means that at the very beginning, the mutual information between $R_+$ and $R_-$ is non-zero and the associated radiation entanglement wedge ($R_+\cup R_-$) is in connected phase.\\
Now, at the early times ($t_b \ll \beta$), we observe that the above given general expression of $I(R_+:R_-)$ can be recast to the following form
\begin{widetext}
\begin{eqnarray}\label{Newqeq4}
	I(R_+:R_-)&\approx&\left(\frac{c}{3}\right)\left[\log\left[\left(\frac{\beta}{2\pi}\right)\sqrt{f(\xi)}\cosh\left(\frac{2\pi r^*(b)}{\beta}\right)\right]-\mathrm{sech}\left(\frac{2\pi r^*(b)}{\beta}\right)-\left(\frac{2\pi^2}{\beta^2}\right)\left\{1+\mathrm{sech}\left(\frac{2\pi r^*(b)}{\beta}\right)\right\}t_b^2\right]~.\nonumber\\
\end{eqnarray}
\end{widetext}
The above expression implies that starting from a finite non-zero value (at $t_b=0$), $I(R_+:R_-)$ decreases with the time-scaling $\sim t_b^2$. On the other hand, at the late times ($t_b \gg \beta$), we obtain the following form of the mutual information
\begin{widetext}
	\begin{eqnarray}\label{Newqeq}
I(R_+:R_-)&\approx&\left(\frac{c}{3}\right)\left[\log\left[\left(\frac{\beta}{2\pi}\right)\sqrt{f(\xi)}\right]-2\cosh\left(\frac{2\pi r^*(b)}{\beta}\right)e^{-\left(\frac{2\pi t_b}{\beta}\right)}\right]~.\nonumber\\		
	\end{eqnarray}
\end{widetext}
This in turn means that at late times ($t_b \gg \beta$), $I(R_+:R_-)$ increases with respect to the observer's time $t_b$.\\
Interestingly, one can note by looking at eq.\eqref{Newqeq4} that there exists a particular value of $t_b$ at which the mutual information will be zero and the entanglement wedge $R_+\cup R_-$ will be disconnected\footnote{As we know mutual information between two subsystems, namely, $A$ and $B$ satisfies the non-negative property, that is, $I(A:B)\geq0$. This means zero is the lowest possible value mutual information can have where the correlation between $A$ and $B$ vanishes.}. With this observation in mind, we give the following proposal.
\begin{widetext}
	\noindent\textbf{Proposal I:} \textit{Starting from a finite, non-zero value (at $t_b=0$), the mutual information between $R_+$ and $R_-$ vanishes at  a particular value of the observer's time ($t_b=t_R$).}
\end{widetext}
We now proceed to compute the explicit expression of $t_R$ by using the general expression given in eq.\eqref{Newqeq3}. The above proposal directly implies the following identity
\begin{eqnarray}\label{eq7}
	I(R_+:R_-)|_{t_b=t_R}&=&0~.
\end{eqnarray}
One can solve the above equation to obtain the value of $t_R$. This reads
\begin{eqnarray}
	t_R=\left(\frac{\beta}{2\pi}\right)\cosh^{-1}\left\{\left(\frac{\frac{\beta}{2\pi}\sqrt{f(\xi)}}{1+\frac{\beta}{2\pi}\sqrt{f(\xi)}}\right)\cosh\left(\frac{2\pi r^*(b)}{\beta}\right)\right\}.\nonumber\\
\end{eqnarray}
\noindent It is to be noted that $t_R$ is much smaller in amplitude in comparison with the time scale $t_b = \beta$ or in other other words, this particular time-scale $t_b=t_R$ resides in the early time domain as $t_R\ll\beta$. We now compute the exact value of $S_{vN}(R_{+}\cup R_{-})$ at this particular time ($t_b=t_R$). This gives us the following value
\begin{eqnarray}\label{eq12}
S_{vN}^{t_{b}=t_{R}}(R_+\cup R_-)&=&\frac{c}{3}\log\left[\frac{\left(\frac{\beta\sqrt{f(\xi)}}{2\pi}\right)^{2}}{1+\frac{\beta\sqrt{f(\xi)}}{2\pi}}\cosh\left(\frac{2\pi r^{*}(b)}{\beta}\right)\right]\nonumber\\
&\approx&\frac{c}{3}\log\left[\frac{\beta}{2\pi}\sqrt{f(\xi)}\right]+\frac{c}{6}\left(\frac{r_{+}}{b}\right)^{2}~.\nonumber\\	
\end{eqnarray}
In the last line we have use the fact that $\frac{\beta\sqrt{f(\xi)}}{2\pi}\gg1$ and $b\gg r_{+}$.\\
We now make some comments. The above given proposal suggests that the mutual correlation ($I(R_+:R_-)$) between matter fields on $R_+$ and $R_-$ is non-zero and finite for the time duration $0\leq t_b<t_R$\footnote{The value of $I(R_+:R_-)$ is maximum at $t_b=0$ and after that it decreases for the range $t_b\leq t_R$, vanishing exactly at $t_b=t_R$. This can be observed from the expression given in eq.\eqref{Newqeq4}.}, representing the fact that the associated entanglement wedge of $R_+\cup R_-$ is in connected phase. Further, at $t_b=t_R$, the mutual information between $R_+$ and $R_-$ vanishes and the entanglement wedge $R_+\cup R_-$ moves to the disconnected phase. Once again we would like to mention that $t_R\ll \beta$. These observations strongly indicate that this time $t_R$
\pagebreak
 is nothing but the Hartman-Maldacena time $t_{HM}$. This observation agrees with the one given in \cite{Grimaldi:2022suv}. Furthermore, if we compute the value of $I(R_+:R_-)$ at $t_b=\beta$ (by substituting $t_b =\beta$ in eq.\eqref{Newqeq3}), we get
\begin{eqnarray}
I(R_+:R_-) &=&\left(\frac{c}{3}\right)\log\left[\left(\frac{\beta}{2\pi}\right)\sqrt{f(\xi)}\cosh\left(\frac{2\pi r^*(b)}{\beta}\right)\right]~.\nonumber\\
\end{eqnarray}
This implies that after $t_b=t_R$, $I(R_+:R_-)$ starts to increase with respect to the observer's time $t_b$. This can be seen from eq.\eqref{Newqeq}.
\section{After Page time scenario: probing the role of $I(B_+:B_-)$}
\begin{figure}[htb]
	\centering
	\includegraphics[scale=0.55]{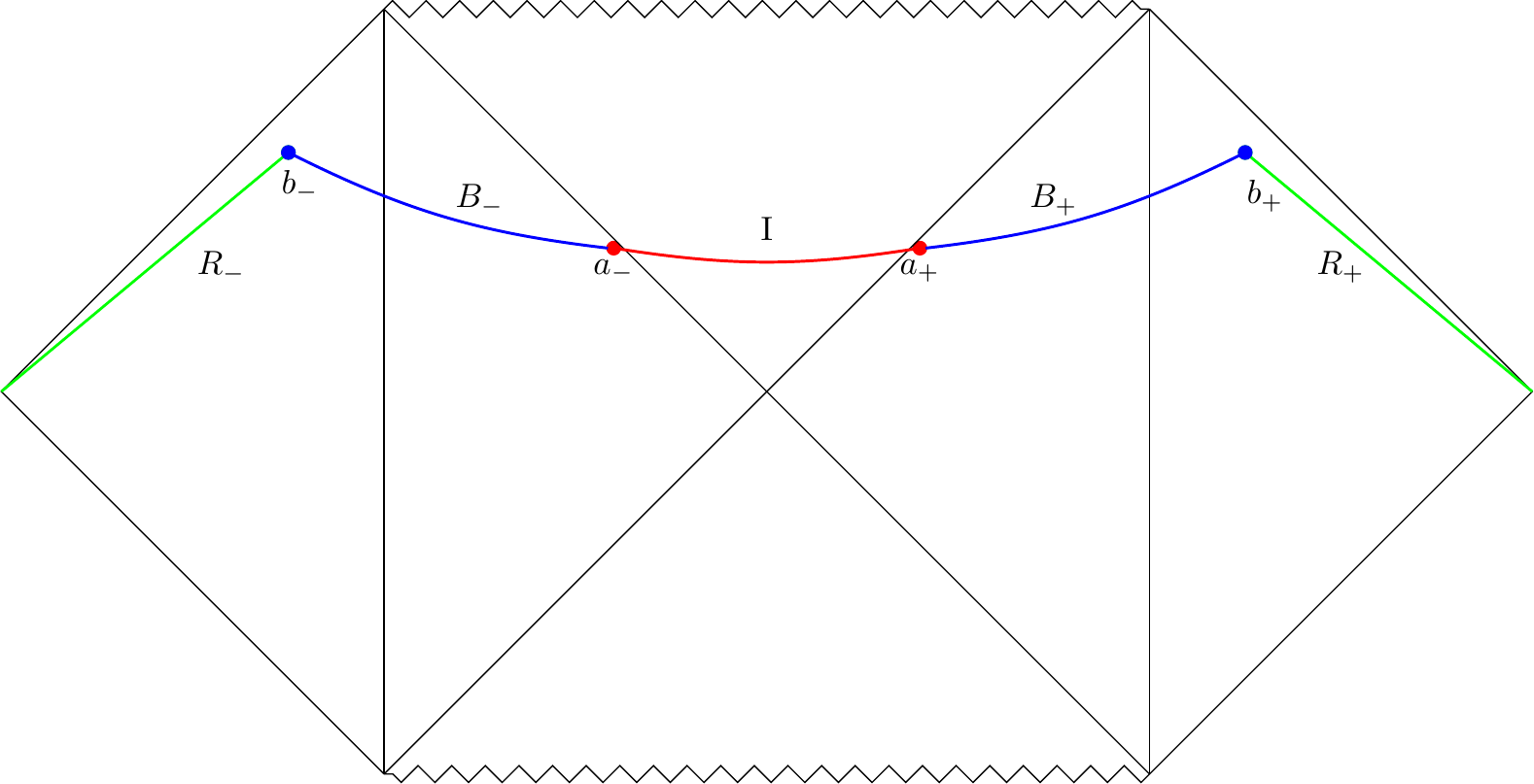}
	\caption{Penrose diagram specifying the island region (in red) with boundaries $a_{\pm}=(\pm t_a,a)$ and the $B_{\pm}$ regions (in blue). The inner boundaries of $B_{\pm}$ regions are $a_{\pm}=(\pm t_a,a)$ and the outer boundaries are $b_{\pm}=(\pm t_b,b)$ \cite{He:2021mst}.}
	\label{fig2}
\end{figure}

We now proceed to discuss the scenario in which the island contribution is dominant. As we have mentioned, just after the Page time $t_p$, the island starts to contribute and one needs to apply the formula given in eq.\eqref{eq1} in order to evaluate the correct fine-grained entropy of radiation. One can observe that the term $S_{vN}(I\cup R)$ satisfies the identity $S_{vN}(I\cup R_+\cup R_-)=S_{vN}(B_+\cup B_-)$. The regions of $B_{\pm}$ can be specified as $(b_{\pm} \rightarrow a_{\pm})$ where $a_{\pm}=(\pm t_a,a)$ are the boundaries of the island. This set up can be represented graphically with the help of a Penrose diagram, as given in Figure \eqref{fig2}. Now as we have mentioned earlier, in this work we are considering $2d$ free CFT as the matter sector. This in turn means that the expression associated to $S_{vN}(B_+ \cup B_-)$ can be evaluated by utilizing the following formula \cite{Calabrese:2009ez}
\begin{widetext}
\begin{eqnarray}\label{eq13}
S_{vN}(B_+\cup B_-)=\left(\frac{c}{3}\right)\log\Big[\frac{d(a_+,a_-)d(b_+,b_-)d(a_+,b_+)d(a_-,b_-)}{d(a_+,b_-)d(a_-,b_+)}\Big]~.
\end{eqnarray}
\end{widetext}
In order to obtain an explicit expression, one needs to substitute the distances, which can be obtained from the black hole metric (given in eq.\eqref{eq4}) and bath metric (eq.\eqref{bath metric}) written in Kruskal coordinates. One can show that for the JT gravity eternal black hole solution, one obtains the following
\begin{eqnarray}\label{eq14}
	S_{vN}(B_+\cup B_-)&=&S_{vN}(B_+)+S_{vN}(B_-)\nonumber\\
	&&+\sim\mathcal{O}(e^{-\frac{2\pi t_{a}}{\beta}}) +\sim\mathcal{O}(e^{-\frac{2\pi t_{b}}{\beta}})
\end{eqnarray}
 where $S_{vN}(B_{\pm})=\left(\frac{c}{3}\right)\log d(b_{\pm},a_{\pm})$. In recent works in this direction, it has been suggested that at the late times ($t_a,t_b\gg\beta$), one obtains the following \cite{Hashimoto:2020cas,Matsuo:2020ypv}
\begin{eqnarray}\label{eq15}
S_{vN}(B_+\cup B_-) \approx S_{vN}(B_+)+ S_{vN} (B_-)	
\end{eqnarray}
if the terms $\mathcal{O}(e^{-\frac{2\pi t_{a,b}}{\beta}})$ are ignored. If one substitutes the above approximate expression in eq.\eqref{eq1} alongwith the correct \textit{area term} and perform the extremization, the end result will be a time-independent expression of $S(R)$, that is $S(R)=2S_{BH}+..$ (for a two-sided eternal black hole). It is to be noted that the above late-time approximation (given in eq.\eqref{eq15}) indirectly gives the hint of vanishing mutual correlation (only in the leading order) between $B_+$ and $B_-$. Keeping this in mind, we give our \textit{after Page time proposal} \cite{Saha:2021ohr}. This reads
\begin{widetext}
\noindent\textbf{Proposal II:}  \textit{The mutual information between the subsystems $B_+$ and $B_-$ vanishes once the island starts to contribute.} 	
\end{widetext}	
We now explain the above proposal from the holographic perspective. We propose that just after the Page time, when the replica wormhole saddle points starts to dominate, the Hawking saddle point (which gives $S(R)\sim t_b$), the entanglement wedge of $B_+\cup B_-$ makes the transition from connected to disconnected phase \cite{Takayanagi:2017knl,Saha:2021kwq,Chowdhury:2021idy} which results in $I(B_+:B_-)=0$. We now apply our proposal. According to our proposal, we need to compute the following
\begin{eqnarray}\label{eq16}
	I(B_+:B_-)&=&0\nonumber\\
S_{vN}(B_+)+S_{vN}(B_-)&=&S_{vN}(B_+\cup B_-)~.	
\end{eqnarray} 
This leads to the following involving the distances
\begin{eqnarray}\label{neweq4}
d^{2}(a_{+},b_{-})&=&d(a_{+},a_{-})d(b_{+},b_{-})~.
\end{eqnarray}
Substituting this condition in $S_{vN}(R_{+}\cup R_{-})$ gives
\begin{eqnarray}\label{neweq5}
S_{vN}(B_{+}\cup B_{-})=\frac{c}{3}\log d^{2}(a_{+},b_{+})~.
\end{eqnarray}
\begin{widetext}
\noindent By using the  gravitational metric given in eq.\eqref{eq3} and the flat metric given in eq.\eqref{bath metric}, we obtain the following explicit expressions 
\begin{eqnarray}
d(a_+,b_+)&=&\sqrt{2F(a)F(\xi,b)e^{\kappa(r^{*}(b)+r^{*}(a))}}\Big[\cosh[\kappa(r^{*}(a)-r^{*}(b))]-\cosh[\kappa(t_{a}-t_{b})]\Big]^{\frac{1}{2}}=d(a_{-},b_{-})\label{neweq6}\\
d(a_-,b_+)&=&\sqrt{2F(a)F(\xi,b)e^{\kappa(r^{*}(b)+r^{*}(a))}}\Big[\cosh[\kappa(r^{*}(a)-r^{*}(b))]+\cosh[\kappa(t_{a}+t_{b})]\Big]^{\frac{1}{2}}=d(a_{+},b_{-})\label{neweq7}\\
d(b_{+},b_{-})&=&2F(\xi,b)e^{\kappa r^{*}(b)}\cosh(\kappa t_{b})\label{neweq8}\\
d(a_{+},a_{-})&=&2F(a)e^{\kappa r^{*}(a)}\cosh(\kappa t_{a})\label{neweq9}~.
\end{eqnarray}
\end{widetext}
Substituting the above expressions in eq.\eqref{neweq4}, we get
\begin{eqnarray}\label{eq17}
t_a-t_b=|r^*(a)-r^*(b)|~.
\end{eqnarray}
The substitution of $t_a$ (in terms of the other quantities) in eq.\eqref{eq13}, leads to the following expression
\begin{eqnarray}\label{eq18}
S_{vN}(B_+\cup B_-)=\frac{c}{3}\log\left[\left(\frac{\beta}{\pi}\right)\sqrt{(\alpha^2-1)(a^2-r_+^2)}\right].\nonumber\\	
\end{eqnarray}
One can obtain the same expression of $S_{vN}(B_{+}\cup B_{-})$ by using the eq.\eqref{neweq5}.
The most striking point about the above expression is that it is independent of time. Now if we substitute the above expression in eq.\eqref{eq1} together with the \textit{area term}, that is $\frac{\mathrm{Area}(\partial I)}{4G_N}=2\times\frac{a}{4G_Nl}$ and extremize with respect to the island parameter $``a"$, we get
\begin{eqnarray}\label{eq19}
	a=r_+ + \left(\frac{2cG_N l}{3}\right)^2\frac{1}{8r_+}+...~.
\end{eqnarray}
Substitution of the above extremized value of $``a"$ leads to the final expression of fine-grained entropy of Hawking radiation. This gives
\begin{eqnarray}\label{eq20}
	S(R)=2S_{BH} -\frac{2c}{3}\log\left(\frac{S_{BH}}{\sqrt{\alpha}}\right)+\frac{\left(\frac{c}{2}\right)^2}{2S_{BH}}+\frac{\left(\frac{c}{3}\right)^3}{32S_{BH}^2}+...\nonumber\\
\end{eqnarray}
where we have used the fact that $\frac{c}{3}\ln(\sqrt{\alpha^{2}-1})\approx- \frac{2c}{3}\ln(\alpha^{-\frac{1}{2}})$, since $\alpha$ is large compared to $1$.
It can be noted from the above expression that it is time independent and contains logarithmic and inverse power law correction terms \cite{Saha:2021ohr}. Revisiting the condition $I(B_+:B_-)=0$ (given in eq.\eqref{eq17}) with the obtained value of ``a" (given in eq.\eqref{eq20}), we get
\begin{eqnarray}\label{eq21}
	t_a-t_b= \left(\frac{\beta}{2\pi}\right)\log\left(S_{BH}\right)=t_{Scr}~
\end{eqnarray}
where $t_{Scr}$ is the \textit{Scrambling time}\cite{Sekino:2008he,Hayden:2007cs}. The remarkable observation made above in turn tells that just after the Page time $t_p$, the replica wormhole saddle points start to dominate and the emergence of island in the black hole interior leads to the disconnected phase of the entanglement wedge $B_+\cup B_-$, characterized by the condition given in eq.\eqref{eq21}. On the other hand, the explicit expression of the Page time is found to be
\begin{eqnarray}\label{eq22}
	t_p = \left(\frac{3\beta}{\pi c}\right) S_{BH}- \left(\frac{\beta}{\pi}\right)\log\left(S_{BH}\right)+\left(\frac{3c}{8}\right) \frac{\beta}{2\pi S_{BH}}+...~.\nonumber\\
\end{eqnarray}
In the above expression, the leading piece is the familiar form of the Page time, where the rest represent the sub-leading corrections to it.
\section{Conclusions}
We now summarize our findings. In this letter we have given two proposals in order to probe the crucial role played by mutual information, in the context of the Page curve. The first proposal is for the before Page time scenario. In this scenario, the Hawking saddle point is the dominating piece which leads to $S(R)\sim t_b$. However, it is only true under the late time ($t_b\gg\beta$) approximation as the early time ($t_b\ll\beta$) growth is found to be $S(R)\sim t_b^2$. With the expression of $S(R)$ in hand, we also compute the von Neumann entropy of matter fields on the intervals $R_+$ and $R_-$. Based upon the obtained results of $S_{vN}(R_+)$, $S_{vN}(R_-)$ and $S_{vN}(R_+\cup R_+)$, the computation of mutual information $I(R_+:R_-)$ has been performed. We then observe the behaviour of $I(R_+:R_-)$ in both early and late times. Motivated by this observation, we propose that at a particular value of $t_b=t_R$, the mutual correlation between the matter fields on $R_+$ and $R_-$ vanishes representing the fact that the associated entanglement wedge $R_+\cup R_-$ becomes disconnected. Further, we note that $t_R\ll\beta$ and $S(R)|_{t_b=t_R}\sim\log\beta$. Based upon these observed facts we then conclude $t_R$ to be the Hartman-Maldacena time $t_{HM}$. Furthermore, we also observe that after $t_b=t_R$, $I(R_+:R_-)$ starts to increase identifying the connected phase of entanglement wedge $R_+\cup R_-$. We now come to our second proposal which we have given for the after Page time scenario. In this scenario, the replica wormholes are the dominating saddle points and the island emerges in the black hole interior. We observe that (following recent works in this direction) one has to consider the late-time approximation (given in eq.\eqref{eq15}) in order to obtain a correct Page curve and further it leads to the extremization condition $t\approx t_b$ \cite{Hashimoto:2020cas}. Motivated by this we propose that the sole working of the island is to disconnect the entanglement wedge of $B_+\cup B_-$, that is, once the island starts to dominate, the mutual information between $B_+$ and $B_-$ vanishes. Remarkably we find that the associated condition is $t_a-t_b=t_{Scr}$ where $t_{Scr}$ is the \textit{Scrambling time} \cite{Sekino:2008he,Hayden:2007cs}. With the help of subadditivity condition of von Neumann entropy we can recast the above observation in the following way. As long $t_a-t_b<t_{Scr}$, the entanglement wedge $B_+\cup B_-$ is in connected phase and once the condition $t_a-t_b=t_{Scr}$ is satisfied, the entanglement wedge of $B_+\cup B_-$ gets disconnected. We also find that this condition of vanishing mutual information leads to a time-independent expression of $S_{vN}(I\cup R)$ and we also find the final expression of $S(R)$ contains universal corrections (logarithmic and inverse power law corrections). This in turn means that our proposals and observations related to mutual information gives strong realization of the concept given in \cite{Grimaldi:2022suv,VanRaamsdonk:2010pw}.\\
To conclude, we propose that in the before Page time scenario,  there is a particular time-scale at which the radiation ($R_+\cup R_-$) entanglement wedge makes transition from connected to disconnected phase and in the after page time scenario, there is a time-scale at which the black hole ($B_+ \cup B_-$) entanglement wedge makes the same transition. Furthermore, the above observations can also be extended to higher dimensional scenarios with $s$-wave approximation in the conformal field theory matter sector \cite{Polchinski:2017dkv,Hemming:2000as}.  
\section{Acknowledgements}
\noindent ARC would like to acknowledge SNBNCBS for fellowship. AS would like to acknowledge the
support by Council of Scientific and Industrial Research (CSIR, Govt. of India) for the Senior Research Fellowship. The authors would like to thank the anonymous referee for very useful comments.
\bibliography{Reference}
\end{document}